\documentclass[conference,10pt]{IEEEtran}
\usepackage{cite}
\usepackage{etoolbox}
\usepackage{mathtools, cuted}
\usepackage{tabu}
\usepackage{array}
\usepackage{comment}

\ifCLASSINFOpdf
\else
\fi

\usepackage{amsmath, amsthm, amssymb, amsfonts}

\usepackage{amsmath}
\usepackage{graphicx}
\usepackage{epstopdf}

\makeatletter
\DeclareRobustCommand\bfseriesitshape{%
\not@math@alphabet\itshapebfseries\relax
\fontseries\bfdefault
\fontshape\itdefault
\selectfont
}
\makeatother

\DeclareTextFontCommand{\textbfit}{\bfseriesitshape}

\begin{document}

%
\title{ On the Asymptotic Throughput of the $k$-th Best Secondary User Selection in Cognitive Radio Systems  }

\author{\IEEEauthorblockN{ Yazan H. Al-Badarneh, Costas N. Georghiades, Mohamed-Slim Alouini}}

\maketitle 

\begin{abstract}
We analyze the asymptotic average and effective throughputs of a multiuser diversity scheme for a secondary multiuser network consisting of multiple secondary users (transmitters) and one secondary receiver. Considering a transmit power adaptation strategy at the secondary users to satisfy the instantaneous interference constraint at the primary receiver, the secondary receiver selects the $k$-th best secondary user for transmission, namely, the one with the $k$-th highest signal-to-noise ratio (SNR). We use extreme value theory to show that the $k$-th highest SNR converges uniformly in distribution to an inverse gamma random variable for a fixed $k$ and large number of secondary users. We use this result to derive closed-form asymptotic expressions for the average and effective throughputs of the $k$-th best secondary user. 

\end{abstract}

\section{Introduction}
Cognitive radio (CR) is an important technology to maximize radio spectrum utilization efficiency\cite{819467}\nocite{4840529}-\cite{1391031}. In CR systems, the secondary network is allowed to share the spectrum allocated to the primary network provided that the interference caused by the secondary transmitter (ST) does not deteriorate the performance of the primary network. Consequently, the challenge is to maintain the interference caused by the ST to the primary receiver (PR) below a pre-determined threshold level. This can be achieved by adapting the ST transmit power that ensures satisfaction of the interference constraint at the PR \cite{4786456}. 

Multiuser diversity is considered an important diversity technique to improve wireless communication systems performance \cite{tse2005fundamentals}. Considering a multiuser network where the users experience independent fading conditions, the basic idea of multiuser diversity is to select the users with the best fading conditions for transmission or reception to obtain a specific performance gain. Multiuser diversity in CR systems has attracted much attention recently. 
 In particular, the ergodic capacity (throughput) of multiuser diversity gain of uplink multiuser underlay CR systems is investigated in \cite{4786488}. In \cite{ekin2012capacity}, the authors analyze the achievable capacity gain of uplink multiuser spectrum-sharing systems over dynamic fading environments. In \cite{li2013capacity}, the outage probability and effective capacity are analyzed for opportunistic spectrum sharing in Rayleigh fading environment. In \cite{khan2015performance}, the authors analyze the outage probability, average symbol error rate (SER) and ergodic capacity of an opportunistic multiuser cognitive network with multiple primary users assuming the channels in the secondary network are independent but not identical Nakagami-$m$ fading. In \cite{7881835}, \cite{AGHAZADEH2018160} the authors analyze the outage probability and average capacity of multiuser diversity in single-input multiple-output (SIMO) spectrum sharing systems.

Related previous work has focused on conventional multiuser diversity in underlay CR systems where the secondary user (SU) with the best channel quality is selected. However, in practical underlay CR systems the SU with the best channel quality may not be available for transmission under given traffic conditions. Consequently for such systems, a more general multiuser diversity scheme that features selection of the SU with the $k$-th best channel quality ($k$-th best SU) is of practical interest. In addition, selecting the best SU is not always beneficial since it prevents other users with good channel quality from transmission or reception although it can achieve maximum diversity gain. One might sacrifice some diversity gain by selecting the second, third or in general the $k$-th best SU to improve the performance from a fairness standpoint.

In this paper, we analyze the average and effective throughputs of a multiuser diversity scheme for secondary multiuser networks under a transmit power adaptation strategy such that the instantaneous interference constraint at the PR is not violated. The SR is equipped with multiple receive antennas where maximal ratio combining (MRC) is employed. Based on the SR-MRC output, the SR selects the $k$-th best SU for transmission. In general, it is hard to find exact and tractable expressions for the average and effective throughputs of the $k$-th best SU. This difficulty is due to the complicated nature of the distribution of the $k$-th highest SNR. Therefore, another approach based on extreme value theory (EVT) is used to analyze the throughput of the $k$-th best SU selection scheme in underlay CR systems. Recently, EVT has been used to derive simple closed form asymptotic expressions for the average and effective throughputs of the $k$-th best link selection for traditional wireless communication systems \cite{8269400}. Our contribution in this paper is to utilize EVT to analyze the average and effective throughputs of the $k$-th best SU selection scheme of underlay CR systems. More specifically,  we show that the SNR of the $k$-th best SU converges uniformly in distribution to an inverse gamma random variable for a fixed $k$ and large number of secondary users. Then, we derive novel closed-form asymptotic expressions for the average and effective throughputs of the $k$-th best SU.

The rest of this paper is organized as follows. In Section II we discuss the system model. In Section III we discuss the asymptotic average and effective throughputs. Section IV includes numerical results and Section V concludes.

\section{System model}
Consider an underlay secondary network consisting of $N$ secondary users (transmitters), each equipped with a single antenna, and a secondary receiver equipped with $M$ receive antennas.  The secondary network is sharing the spectrum of a primary network with one PT and one PR. The PT and PR are equipped with a single antenna each. Let $g_{i}$ and $h_{i,j}$ denote the channel gain from the $i$-th SU  to the PR and the $j$-th receive antenna of the SR, respectively. The channel gains $g_{i}$ and $h_{i,j}$ are assumed to be independent Rayleigh distributed random variables. Consequently, the channel power gains $|g_{i}|^{2}$ and $|h_{i,j}|^{2}$ have  probability density functions (PDFs) $g(x) =  \lambda e^{-\lambda  x} u(x) $ and $h(x) = \eta e^{- \eta x}u(x)$, respectively, where $u(x)$ is the unit step function and the parameters $\lambda$ and $\eta$ are the fading parameters. The channel power gains in the secondary system $|h_{i,j}|^{2}$ are assumed to be independent and identically distributed (i.i.d)  for $i=1, 2, ..., N$ and $j=1, 2, ..., M$. 

With a perfect knowledge of $|g_{i}|^{2}$, we consider a continuous transmit power adaptation strategy at each SU to control its interference to the PR such that the instantaneous transmit power of the $i$-th SU is $P_{i}= \frac{Q}{|g_{i}|^{2}}$, where $Q$ is the maximum tolerable interference level at the PR. It should be noted that this paper focuses on a power adaptation strategy with unlimited transmit power as in \cite{6134707}, \cite{7890994}. An efficient power adaptation strategy with peak transmit power constraint is left for future work.

Assuming that MRC is employed at the SR, the instantaneous SNR at the SR-MRC output is given by
\begin{gather} \label{eq:1}
Z_{i}= \frac{Q}{|g_{i}|^{2}} \frac{\gamma_{i}}{N_{0}} . 
\end{gather} 
where $N_{0}$ is the common noise variance at the SR-MRC output and $\gamma_{i}= \sum_{j=1}^{M} |h_{i,j}|^{2}$. The random variable $\gamma_{i}$ represents a sum of i.i.d exponential random variables, therefore, its PDF is given by  
\begin{equation} \label{eq:2}
f_{\gamma_{i}}(x)={ \frac{\eta^{M} x^{M-1}}{ \Gamma(M)} e^{-\eta x}} u(x). 
\end{equation}

Based on the SR-MRC output, we sort the random variables $Z_{i}$ in an increasing order denoted as  $Z_{(1)} \leq Z_{(2)}.... \leq Z_{(N-k+1)} \leq .... \leq Z_{(N)}$, such that the SR selects the $k$-th best SU that leads to the $k$-th highest SNR, $Z_{(N-k+1)}$. According to \cite{david2003order}, the PDF of $Z_{\left(N-k+1\right)}$ can be expressed in terms of the PDF, $f(z)$, and cumulative distribution function (CDF), $F(z)$, of $Z_{i}$ as 
\begin{equation}\label{eq:3}
f_{Z_{\left(N-k+1\right)}}(z)=k \binom{N}{k} f(z) F(z)^{N-k} \left(1-F(z) \right)^{k-1}.  
\end{equation}
Noting that the random variable $Z_{i}$ represents a ratio of two gamma distributions, its CDF is given by \cite{6924725} 
\begin{equation}\label{eq:4}
F(z)=\left( \frac{\rho z}{\frac{\lambda}{\eta}  +\rho z}\right)^{M} u(z), 
\end{equation}
where $\rho=\frac{N_{0}}{Q}$.  Considering the $k$-th best SU selection, the average (ergodic) throughput of the selected SU, $ \overline{R}_{k,N}$, can be evaluated as 
\begin{equation}\label{eq:6}
\begin{split}
 \overline{R}_{k,N}&=B E\left[\log_{2}(1+Z_{\left(N-k+1\right)})\right]\\
&=B \int_{0}^{\infty} \log_{2}(1+z) f_{Z_{\left(N-k+1\right)}}(z) dz. 
\end{split}
\end{equation}
where $B$ is the system bandwidth and $E[\cdot]$ denotes expectation. 

Assuming a block fading channel, the effective throughput that can be supported by a wireless system under a statistical QoS constraint described by the delay QoS exponent $\theta$ as \cite{1210731}
\begin{eqnarray} \label{eq:7}
\alpha (\theta) =-\frac{1}{\theta T} \log \left( E\left[ e^{-\theta T R }\right] \right),  \ \theta> 0,
\end{eqnarray}
where $R$ is a random variable which represents the instantaneous throughput during a single block and $T$ is the block length. $\theta=0$ implies there is no delay constraint and the effective throughput is then the average throughput of the corresponding wireless channel.  

Considering the $k$-th best SU selection, the effective throughput of the selected SU, $\alpha(\theta, k, N)$, can be expressed as \cite{6006584}
\begin{gather} \label{eq:8}
\begin{split}
\alpha(\theta, k, N)&=-\frac{1}{A} \log_{2} \left(  E\left[ \left( 1+Z_{(N-k+1)} \right)^{-A} \right] \right),  
\end{split}
\end{gather}
where $A=\theta TB/ \ln(2)$. 
In general, it is difficult to obtain exact expressions for $  \overline{R}_{k,N}$ and $\alpha(\theta, k, N)$. Therefore, in what follows we consider extreme value theory to derive closed-form asymptotic expressions for the average and effective throughputs of $k$-th best SU. 

\section{ Asymptotic Throughput Analysis} 
In this section, we derive the limiting distribution of $Z_{\left(N-k+1\right)}$ in Proposition 1 below. Based on this result we will analyze the average and effective throughputs of the $k$-th best SU.

\subsection{ The Limiting Distribution of  $Z_{\left(N-k+1\right)}$} 

\noindent{\textbf{Proposition 1:}}  Let $Z_{(N-k+1)}$ denote the $k$-th largest order statistic of $N$ i.i.d. random variables with a common CDF of $F(z)$, as expressed in (\ref{eq:4}), then for a fixed $k$ and $N \to \infty$, $\frac{ Z_{(N-k+1)}- a}{b}$ converges in distribution to a random variable $Z$ with CDF $G^{(k)}(z)$, 
which can be characterized by an inverse gamma distribution as
\begin{eqnarray}\label{eq:11D}
\begin{split}
G^{(k)}(z)=\frac{\Gamma  \left( k,{\frac {1}{z}} \right)}{(k-1)!} u(z) ,  
\end{split}
\end{eqnarray}
where $a=0$, $b=\frac{ \beta \lambda}{P_{M} \left(  \left(1-\frac{1}{N}\right)^{-\frac{1}{m}} -1 \right)} >0$ and $\Gamma(s,x)= \int_{x}^{\infty} u^{s-1} e^{-u} du$  is the upper incomplete gamma function \cite{xxx}.
Furthermore, the PDF of $Z$, $f^{(k)}(z)$, can be obtained as
\begin{eqnarray}\label{eq:12D}
f^{(k)}(z)= \frac{e^{- z^{-1}}}{z^{k+1} (k-1)!} u(z).  
\end{eqnarray} 

\noindent \textit{Proof}: We first obtain the limiting distribution of $Z_{(N)}$, which denotes the first largest order statistic of $N$ i.i.d. random variables. From  Proposition 2 of  \cite{6924725}, $\frac{ Z_{(N)}- a}{b}$ converges in distribution to a unit Fr\'echet distribution i.e.,
\begin{eqnarray}\label{eq:13D}
G(z)= e^{-z^{-1}} u(z),
\end{eqnarray}
 where $a=0$ and $b = { F^{-1}\left(1-\frac{1}{N}\right)}=\frac{ \beta \lambda}{P_{M} \left(  \left(1-\frac{1}{N}\right)^{-\frac{1}{m}} -1 \right)}$. Making use of Proposition 1 of \cite{8269400} with $G(z)$  as in (\ref{eq:13D}), it follows that for a fixed $k$ and $N \to \infty$, the sequence $\frac{ Z_{(N-k+1)}}{b}$ converges in distribution to a random variable $Z$ with CDF of $G^{(k)}(z)$, which can be expressed in terms of $G(z)$  as
\begin{eqnarray}\label{eq:14D}
\begin{split}
G^{(k)}(z)&=G(z) \sum_{j=0}^{k-1} \frac{\left[ -\log \left( G(z) \right) \right]^{j}}{j !}\\
&= e^{-z^{-1}}\sum_{j=0}^{k-1} \frac{(z^{-1})^j }{j!} u(z).\\
\end{split}
\end{eqnarray}
Using the fact that $\Gamma(k,x)= (k-1)! \ e^{-x}\sum_{j=0}^{k-1} \frac{x^j }{j!}$ for an integer $k$, $G^{(k)}(z)$ can be finally expressed as in (\ref{eq:11D}). By differentiating (\ref{eq:11D}) we obtain (\ref{eq:12D}).

Note that  Proposition 1 of \cite{8269400} can be applied for different CDF functions. In this paper we focus on the case when $G(z)$ represents a Fr\'echet CDF. In this case $Z_{(N-k+1)}$ has a limiting distribution of inverse gamma as shown in (\ref{eq:11D}). This is different from what was obtained in \cite{8269400}, where Proposition 1 of \cite{8269400} was applied for the case when $G(z)$ represents Gumbel CDF and thus $Z_{(N-k+1)}$ has a limiting distribution of $\textit{Log-Gamma}$.

\subsection{ Asymptotic average throughput}
Using Proposition 1, we derive the average throughput for the $k$-th best SU, $ \overline{R}_{k,N}$, in the following proposition. \\

\noindent{\textbf{Proposition 2:}} For a fixed $k$ and $N \to \infty$, the average throughput of the $k$-th best SU can be approximated as
\begin{gather} \label{eq:14}
\begin{split}
\frac{\overline{R}_{k,N}}{B}\approx &
\frac{ \ln(b) - \psi(k)}{\ln(2)} +\frac{1}{\ln(2)} \sum_{\mu=0}^{k-1} \frac{ 1}{ (k-\mu-1)!} \times \\
& \left[ {\left(-1\right)^{k-\mu-2}  b^{k-\mu-1} e^{b}  E_{i}(-b) } \right. \\ 
&\left. +{ \sum_{v=1}^{k-\mu-1} (v-1)! (-b)^{k-\mu-1-v}}  \right], 
\end{split}
\end{gather}
in \text{bit/s/Hz}, where  $E_{i}(x)=-\int_{-x}^{\infty} \frac{e^{-y}}{y} dy$ is the exponential integral function and $\psi(x)$ is the digamma function. 

\noindent \textit{Proof}: 
Invoking (\ref{eq:6}) we have
\begin{gather}\label{eq:16}
\begin{split}
\frac{\overline{R}_{k,N}}{B}&=\frac{1}{\ln(2)} E\left[\ln\left(1+Z_{(N-k+1)} \right) \right].\\
\end{split}
\end{gather}
From Proposition 1, the CDF of $\frac{ Z_{(N-k+1)}}{ b}$ approaches the CDF $Z$ for a fixed $k$ and $N \to \infty$, where the CDF of $Z$ is as in (\ref{eq:11D}). Or equivalently, the PDF of $ Z_{(N-k+1)}$ can be approximated by the PDF of $  b Z$ for a fixed $k$ and $N \to \infty$, where  the PDF of $Z$ is as in (\ref{eq:12D}). Then for a fixed $k$ and $N \to \infty$, the average throughput can be approximated as 
\begin{gather}\label{eq:17}
\begin{split}
\frac{\overline{R}_{k,N}}{B} &\approx\frac{1}{\ln(2)} E\left[\ln\left(1+ b Z \right) \right]\\
& = \int_{0}^{\infty} \frac{ \ln(1+b z)}{\ln(2)} \frac{ e^{- z^{-1}} }{ z^{k+1} (k-1)!} dz. 
\end{split}
\end{gather}
Using change of variables $u= z ^{-1}$, then we can write 
\begin{gather}\label{eq:18}
\begin{split}
\frac{\overline{R}_{k,N}}{B} \approx & 
\underbrace{\int_{0}^{\infty} \frac{\ln(b+u) e^{-u} u^{k-1}}{\ln(2) (k-1)!} du}_\text{$I_{1}$}\\
&- \underbrace{ \int_{0}^{\infty} \frac{\ln(u) e^{- u} u^{k-1} }{\ln(2) (k-1)!} du}_\text{$I_{2}$} . 
\end{split}
\end{gather}
Using Eq. (4.352, 1) of  \cite{xxx}, $I_{2}$ can be expressed as 
\begin{eqnarray}\label{eq:19}
\begin{split}
I_{2}= \frac{\psi(k)}{\ln(2)}. 
\end{split}
\end{eqnarray}
To evaluate $I_{1}$, we use Eq. (4.337, 5) of  \cite{xxx}. After some basic algebraic manipulation, $I_{1}$ can be expressed as 
\begin{gather} \label{eq:20}
\begin{split}
I_{1}= &
\frac{ \ln(b)}{\ln(2)} +\frac{1}{\ln(2)} \sum_{\mu=0}^{k-1} \frac{ 1 }{ (k-\mu-1)!} \times \\
& \left[ {\left(-1\right)^{k-\mu-2}  b^{k-\mu-1} e^{b}  E_{i}(-b) } \right. \\ 
&\left. +{ \sum_{v=1}^{k-\mu-1} (v-1)! (-b)^{k-\mu-1-v}}  \right].  
\end{split}
\end{gather}
Combining (\ref{eq:19}) and (\ref{eq:20}) with (\ref{eq:18}), the average throughput is as expressed in (\ref{eq:14}). As a special case, if $k=1$ in (\ref{eq:14}), $\psi(1)=-\gamma$ (Euler's constant); thus
\begin{gather}\label{eq:23}
\begin{split}
\frac{\overline{R}_{1,N}}{B} \approx \frac{\ln(b) +\gamma- e^{b} E_{i}(-b) }{\ln(2)}.
\end{split}
\end{gather}

\subsection{Asymptotic Effective Throughput}
Using Proposition 1, we analyze the effective throughput of the $k$-th best SU, $\alpha (\theta, k, N)$, in the following proposition.  \\

\noindent{\textbf{Proposition 3:}}
 The effective throughput of the $k$-th best SU can be approximated as 
\begin{gather} \label{eq:24}
\begin{split}
&\alpha (\theta, k, N) \approx  -\frac{1}{A} \log_{2} \left( \frac{b^{k} U\left(A+k; k+1;b\right) \Gamma\left(A+k\right)}{(k-1)!} \right), \\
\end{split}
\end{gather}
for a fixed $k$, $ \theta >0$ and $ N \to \infty$, where $ \Gamma( \cdot )$ is the gamma function and $\textit{U}\left( a;b;z \right)=\frac{1}{\Gamma(a)} \int_{0}^{\infty} e^{-zt} t^{a-1} (1+t)^{b-a-1}dt$, $a > 0$ is the Tricomi hypergeometric function. 

\noindent{\textit{Proof:}} 

As we mentioned earlier, the PDF of $ Z_{(N-k+1)}$ can be approximated by the PDF of $  b Z$ for a fixed $k$ and $N \to \infty$, where  the PDF of $Z$ is as in (\ref{eq:12D}). Then for a fixed $k$ and $N \to \infty$, $ E\left[ \left( 1+ Z_{(N-k+1)} \right)^{-A}\right]$ in (\ref{eq:8}) can be approximated as  
\begin{gather} \label{eq:25}
\begin{split}
 E\left[ \left( 1+ Z_{(N-k+1)} \right)^{-A} \right] &\approx E\left[ \left( 1+ b Z \right)^{-A} \right] \\
&=\underbrace{ \int_{0}^{\infty}   \frac{\left( 1+ b z \right)^{-A}  e^{-z^{-1}} }{ z^{k+1} (k-1)!} dz.}_\text{$I_{3}$}
\end{split}
\end{gather}
Using change of variables $u=(b z)^{-1}$ and with the help of Eq. (39) of \cite{1576535}, $I_{3}$ can be finally expressed as   
\begin{gather} \label{eq:26}
\begin{split}
I_{3} &= \int_{0}^{\infty}  \frac{b^{k} u^{A+k-1}   e^{-b u} }{ (1+u)^{A} (k-1)!} du\\
&=\frac{b^{k} U\left(A+k; k+1;b\right) \Gamma\left(A+k\right)}{(k-1)!} . 
\end{split}
\end{gather}
Substituting  (\ref{eq:26}) in  (\ref{eq:8}), we obtain  (\ref{eq:24}). 

\section{ NUMERICAL RESULTS } 
In this section, we numerically illustrate and verify the derived  asymptotic results in the previous section. In Fig. 1, we plot the average throughput  as a function of the number of secondary users, $N$, for $M=2$ and different values of $k$. We use Monte Carlo simulations to validate the obtained asymptotic expression for the average throughput. Compared to the simulations, we observe that the asymptotic expression is accurate for large $N$ relative to $k$. However, if $N$ is close enough to $k$, for example when $N=5$ and $k=3$, the  asymptotic expression tends to be less accurate. This is due to the fact that the asymptotic behavior of the $k$ highest SNR holds for large $N$ and fixed $k$ as discussed in the previous section. Therefore, it is expected that the asymptotic average throughput will be less accurate if $N$ is close enough to $k$. 

\begin{figure}[h] \label{fig:1}
\begin{center}
\includegraphics[width=1\columnwidth]{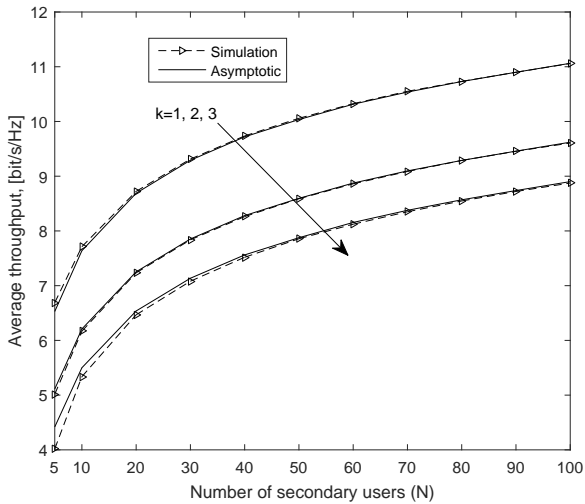}
\caption{Average throughput of the $k$-th best SU versus the number of secondary users, $N$, for $M=2$, $\rho=1$, $\lambda=2$, $\eta=\frac{1}{3}$ .} 
\end{center}
\end{figure}
\begin{figure}[h] \label{fig:2}
\begin{center}
\includegraphics[width=1\columnwidth]{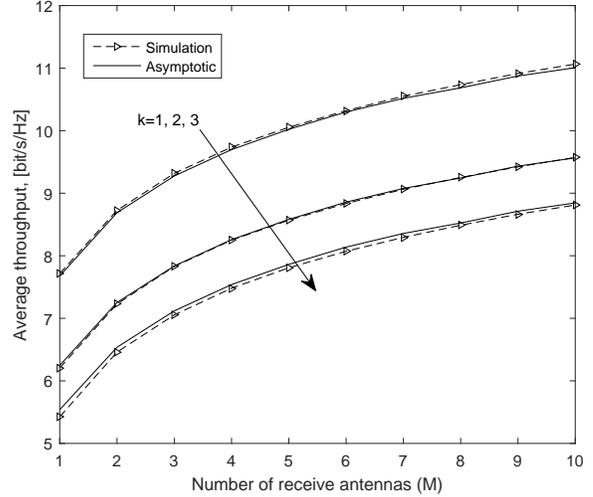}
\caption{Average throughput of the $k$-th best SU versus the number of receive antennas, $M$, for $N=20$, $\rho=1$, $\lambda=2$, $\eta=\frac{1}{3}$.} 
\end{center}
\end{figure}

In Fig. 2 the average throughput is plotted against the number of receive antennas, $M$, for $N=20$  and different values of $k$. In Fig. 3, We plot the effective throughput as a function of the number of secondary users, $N$, for $M=1$, $k=1,2$ and different values of delay exponent $A$. We verify the accuracy of the asymptotic effective throughput using simulations.  We also observe that the asymptotic  effective throughput is accurate for large $N$ and less accurate as $N$ gets closer to $k$ as previously observed for the average throughput.  In Fig. 4, we plot the effective throughput at $A=0$ (average throughput) and at $A=1$ as a function of the maximum tolerable interference level, $Q$, for $N=50$, $M=3$ and $k=1,2$. 
\begin{figure}[h] \label{fig:3}
\begin{center}
\includegraphics[width=1\columnwidth]{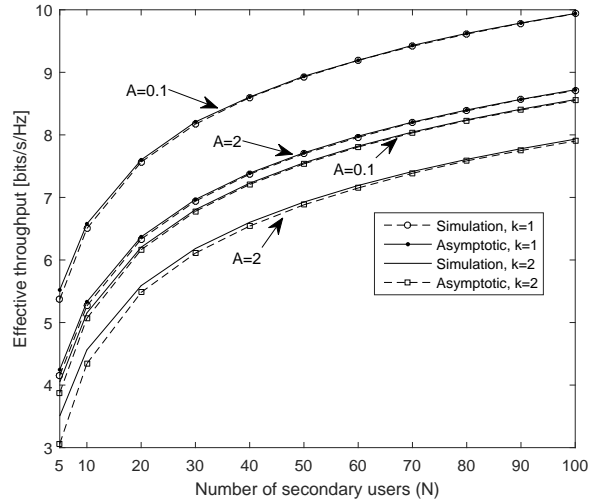}
\caption{Effective throughput of the $k$-th best SU versus the number of secondary users, $N$, for $A=0.1$, 2,  $M=1$, $\rho=1$, $\lambda=2$, $\eta=\frac{1}{3}$.} 
\end{center}
\end{figure}

\begin{figure}[h] \label{fig:4}
\begin{center}
\includegraphics[width=1\columnwidth]{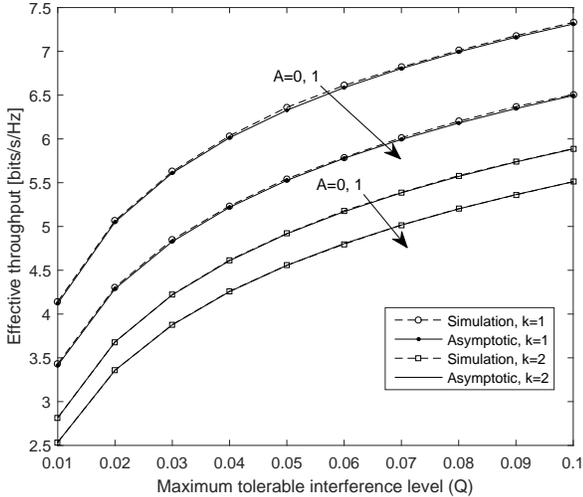}
\caption{Effective throughput of the $k$-th best SU versus the maximum tolerable interference level, $Q$,  for $A=0, 1$,  $M=3$, $N=50$, $\lambda=2$, $\eta=\frac{1}{3}$ and $N_{0}=0 \ dB$.} 
\end{center}
\end{figure}

\section{Conclusion}
We considered a multiuser diversity scheme for cognitive radio system with transmit power adaptation strategy which ensures the satisfaction of the instantaneous interference constraint at the PR. Assuming a large number of secondary users and the SR selects the $k$-th best SU for transmission, we showed that the $k$-th highest SNR converges in distribution to an inverse gamma random variable. We used this result to derive novel closed-form asymptotic expressions for the average and effective throughputs of the $k$-th best SU. We verified the accuracy of the derived expressions through Monte Carlo simulations. 

\section*{Acknowledgement} 
This publication was made possible by the NPRP award [NPRP 8-648-2-273] from the Qatar National Research Fund (a member of The Qatar Foundation). The statements made herein are solely the responsibility of the authors.
\bibliographystyle{IEEEtran}
\bibliography{yvlc}
\end{document}